# Room- and low-temperature magnetic parameters of $Y_3AlFe_4O_{12}$ garnets

Vladyslav Borynskyi[1*], Dariia Popadiuk[1,2], Anatolii Kravets[1,2], Yuliia Shlapa[3], Serhii Solopan[3], Vladislav Korenivski[2], Anatolii Belous[3], Alexandr Tovstolytkin[1]

[1]*Institute of Magnetism of the NAS of Ukraine and MES of Ukraine,*
*36-b, Akad. Vernadskogo blvd., Kyiv 03142, Ukraine*
[2]*Nanostructure Physics, Royal Institute of Technology,*
*12, Hannes Alfvénsväg, Stockholm 10691, Sweden*
[3]*V.I. Vernadsky Institute of General and Inorganic Chemistry of the NAS of Ukraine,*
*32/34, Palladina ave., Kyiv 03142, Ukraine*

**Abstract**

Magnetic properties of $Y_3AlFe_4O_{12}$ garnet ceramics have been studied over a wide range of temperature (3 – 370 K) and magnetic fields (up to 2 kOe). Effects of varying the temperature on some of the application-specific magnetic characteristics have been analyzed in detail. With the decrease in temperature, the effective anisotropy constant, $K_{Eff}$, is found to sharply rise below ~150 K, while the saturation magnetization, $M_s$, changes only slightly (<20 % within 3 – 250 K). The exchange stiffness, $A_{ex}$, and spin wave stiffness, $D$, parameters mirror the relatively smooth behavior of the magnetization at low temperatures but display a rapid drop when $T$ approaches $T_C \approx 436$ K. The temperature dependence of the domain wall thickness and the critical single-domain size correlate with the corresponding values for pure $Y_3Fe_5O_{12}$ (YIG) and are in agreement with the theoretical estimates. We conclude by discussing the ways of how to use the obtained results for tailoring the magnetic properties of YIG-based materials for technological applications.

**Key words:** Al-doped yttrium iron garnets, magnetic parameters, ferrimagnetic ordering, temperature effects, spin wave stiffness, domain structure.

## 1. Introduction

Rapid development of information and communication technologies, and a pronounced trend towards the nanoscale fabrication technique impose increasingly stringent requirements on the materials that form the basis of the elements of modern information systems [1–3]. In the fields of microwave electronics, spintronics, magnonics, optoelectronics, etc. the materials based on iron-yttrium garnet $Y_3Fe_5O_{12}$ (YIG) are widely used owing to their low magnetic damping, controllable saturation magnetization, high electrical resistivity, and large magneto-optical Faraday rotation [4–6]. The enhanced attention to this class of materials has resulted in reliable establishment of the approaches to tailor their magnetic properties (magnetization, effective magnetic anisotropy, magnetostriction constants, and hysteresis parameters) to those suitable for advanced applications [7,8]. At the same time, the adjustment of the corresponding magnetic properties for particular frequency or temperature ranges often remains a tough challenge [9–12].

YIG's crystal structure mirrors that of a natural garnet whose crystals belong to the cubic space group *Ia*-3*d* [13]. Each crystal cell of YIG comprises 8 formula units, consisting of 24 $Y^{3+}$ cations in dodecahedral (*c*) sites, 16 $Fe^{3+}$ cations in octahedral (*a*) sites, 24 $Fe^{3+}$ cations in tetrahedral (*d*) sites, and 96 $O^{2-}$ ions. In pure YIG, the strongest magnetic interactions are attributed to inter-sublattice exchange, specifically to super-exchange interactions between $Fe^{3+}$ ions of the octahedral and tetrahedral sites, mediated by $O^{2-}$ ions [1].

---

* Corresponding author: vladislav.borinskiy@gmail.com

The complex lattice and magnetic structure of YIG provide the possibility for efficient control of various properties [14]. A promising way to purposefully modify the properties of YIG includes doping and substituting of Fe ions at the *a*-site or *d*-site mainly with magnetic ions such as $Zr^{4+}$ [15–17], $Co^{3+}$ [18], $Mn^{3+}$ [19], $Ni^{2+}$ [20], and others. Such kind of ion substitutions strongly impacts the magnetic properties of YIG. Additionally, extensive research has been aimed at understanding the substitution of $Fe^{3+}$ ions by non-magnetic ones. In recent years, Fe ions substitution by $Ca^{2+}$ [12], $Bi^{3+}$ [21–23], $Al^{3+}$ [24], $In^{3+}$ [25,26] have been studied to modify YIG's magnetic and dielectric properties, and leading to a series of new materials for various potential applications [4,14].

Recently, increased scientific attention has been drawn to the poly- and nanocrystalline ferrites-garnets, in which iron ions are partially substituted by aluminum ones [27–31]. The $Al^{3+}$ ions have been shown to prefer tetrahedral sites for relatively small concentrations, while the distribution over tetrahedral and octahedral sites may occur for higher concentrations [30]. Huang *et al*. [19] reported improved magnetic properties of Al and Mn substituted YIG for phase shifter applications. Other advantages of the Al-doped YIG ferrites-garnets include low dielectric and magnetic losses, reduced coercivity and slightly increased dielectric constant that is necessary for a number of microwave devices [32–35].

Over the last few decades, more and more attention has been paid to the applications working at cryogenic temperatures [9,10,36,37]. The efficient use of superconductors in ferrite-based applications has been successfully demonstrated [38]. Miniature devices with superconducting microstrip circuits can become attractive replacements for bulky room-temperature applications, while offering dramatically improved figures of merit [10,36,39,40]. Unfortunately, the compositions and parameters of the conventional garnet ferrites have not been yet optimized for cryogenic temperatures. For this reason, more attention should be paid to the studies of the properties of YIG-based garnets at low temperatures.

In our recent papers, a new technological way was suggested to fabricate high-quality YIG-based nanomaterials [24,41]. The method we suggest ensures high chemical homogeneity of the samples and in contrast to other methods, it is relatively cheap and provides high productivity. It has been shown that Al-doped YIG samples fabricated according to the new synthesis procedure display the physical-chemical properties which are comparable to or exceeding those inherent in the samples fabricated by a traditional synthesis. Room-temperature magnetic properties of the Al-doped YIG samples fabricated with the use of this procedure were reported in Refs. [24,41]. At the same time, low-temperature magnetic parameters of these samples have been studied poorly.

<u>The aim of this work</u> is to carry out magnetostatic measurements on $Y_3AlFe_4O_{12}$ (YAIG) ceramic samples and extract temperature-dependent intrinsic parameters (saturation magnetization, effective anisotropy constant, spin-wave stiffness coefficient, domain wall thickness) governing static and dynamic magnetic behavior of Al-doped YIG garnets, especially at cryogenic temperatures.

## 2. Experimental details

The ceramic YAIG samples were obtained by sintering of $Y_3AlFe_4O_{12}$ nanopowder synthesized via co-precipitation in aqueous solutions as described in detail in [24,41].

Briefly, aqueous solutions of yttrium, aluminum and iron (III) nitrates were used as starting reagents; the concentration of each salt was 1.5 M. 10 M solutions of NaOH or $NH_4OH$ were used as a precipitator. To obtain $Y_3AlFe_4O_{12}$ precursor, metal hydroxides were precipitated at maintained pH values by means of precipitation of the mixture of $Fe(OH)_3$ and $Al(OH)_3$ at pH 4.0-4.5 followed by the precipitation of $Y(OH)_3$ at pH 8.8-8.9 using $NH_4OH$ solution.

The preliminary synthesized $Y_3AlFe_4O_{12}$ powders subjected to the heat treatment at 800 °C for 2 hours in the air were used for the preparation of the ceramic samples. The products obtained after heat treatment were ground with water in the ball-mill with metallic balls for 4 hours. 15%-solution of the binding component and 3% of aqueous solution of polyvinyl alcohol were added to the powder and corresponding tablets were pressed under the pressure of 2 ton/cm$^2$. The prepared tablets were dried at 70-90 °C, warmed up in the muffle furnace with the heating rate of 5 °C/min to 1350 °C and kept at 1350-1400 °C for 2 hours.

The crystallographic properties of the synthesized $Y_3AlFe_4O_{12}$ garnets were studied via the X-ray diffraction method (XRD) using a DRON-4 diffractometer equipped with a CuKα radiation tube. The morphology of the samples was analyzed via electron microscopy facilities using JEM 1230/JEM 1400 transmission electron microscope (Jeol, Japan) and FEG-SEM Nova Nanosem 230 FEI / SEC miniSEM SNE4500 MB scanning electron microscope, respectively.

The samples for magnetization studies were cut in the form of rectangular parallelepipeds with dimensions $5 \times 3 \times 1$ mm$^3$. Magnetic field was applied parallel to the largest edge of the parallelepiped.

Magnetization measurements were carried out using PPMS DynaCool (Quantum Design Inc.) equipped with a VSM magnetometer. Experimental hysteresis data for a single loop were recorded at a constant temperature with sequential forward and backward field sweep from +2500 Oe to –2500 Oe. After each cycle, the magnetic field was reduced to zero and the sample was heated to a next temperature point. A series of hysteresis loops were obtained at different temperatures in the range of 3 – 370 K.

## 3. Experimental results
### 3.1. Characterization of the samples

According to the XRD data, the synthesized ceramic materials are single-phase (Fig. 1(a)). The crystals belong to the *Ia-3d* cubic space group with a lattice constant $a = 12.321(5)$ Å and volume $V = 1870.8(4)$ Å$^3$. The value of the lattice constant is smaller than that of the pure YIG (12.376 Å) and close to the values reported for $Y_3AlFe_4O_{12}$ in Refs. [28,32].

The samples are visually characterized by the low porosity. The ratio of the measured (4.96 g/cm$^3$) to the calculated (5.03 g/cm$^3$) density is 98.6 %. The average size of the grains in the ceramics is approx. 10 μm (see Fig. 1(b)).

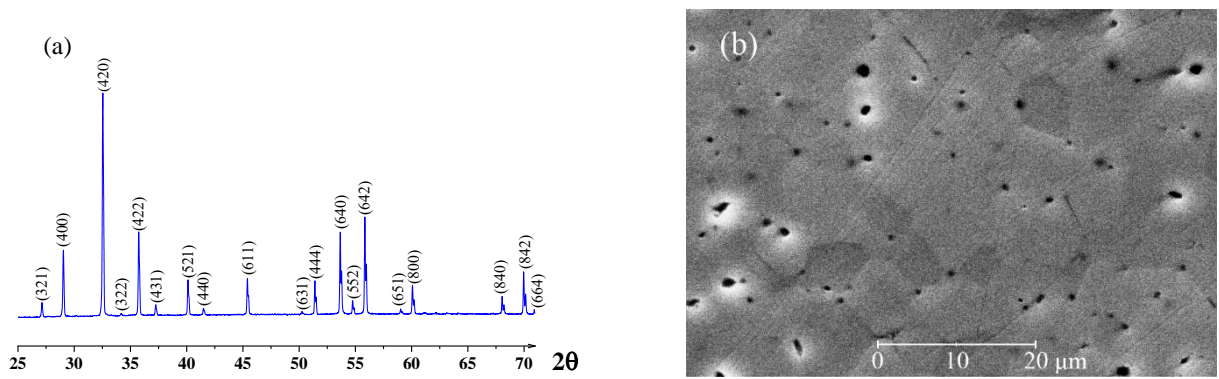

Fig. 1. XRD patterns (a) and the representative SEM image of the polished surface (b) for the ceramic YAIG garnet ferrite under study.

### 3.2. Magnetic properties

Fig. 2(a) shows magnetic hysteresis curves $M(H)$ at selected temperatures measured for the YAIG ceramic sample under study. All curves are characterized by relatively low remanence and coercivity, as is usually observed in soft ferrimagnetic materials (for a better view, Fig. 2(b) depicts the central area of the same loops with increased scale). The shape of the loops can be formally divided into two distinct parts: the low-field part (approx. 0 – 100 Oe), where the net magnetization $M$ rapidly builds up, and the high-field part ($H \gtrsim 300$ Oe) – the so-called approach to a saturation region, wherein the rate of the magnetization change tends to zero. For the YAIG sample under study, the loops show gradual transition between these two regions at all temperatures. This transition unfolds in the intermediate fields of about 100 – 300 Oe where $M$ vs $H$ dependence has a distinctive curvature, indicating the presence of non-negligible magnetic anisotropy which is thought to be randomly distributed between the crystallites [42–44].

The loop at 360 K has narrow hysteresis with $M$ approaching 6.99 emu/g at the external field of 1 kOe, where the magnetization has already reached the plateau. Upon cooling the sample down to 207 K, the first prominent change appears – the magnetization undergoes a significant increase to 11.1 emu/g, which is accompanied by only a small increase in the coercive force $H_c$, from ~2 Oe at 360 K to ~5 Oe at 207 K. Upon further decrease in temperature, the magnetization remains almost unchanged. At the same time, a rapid increase in the coercive field values, up to ~60 Oe at 3 K, is observed.

Temperature dependencies of remnant magnetization $M_r(T)$ and squareness parameter SQ of the YAIG ceramics hysteresis, defined as a ratio $M_r/M_s$, are depicted in Fig. 2(c) as circles and square dots, respectively. Preserving negligibly small values (<1 emu/g) from 360 K down to ~150 K, the $M_r(T)$ dependence exhibits a sharp upturn on further cooling the sample to lower temperatures. The highest value of $M_r$, observed at 10 K, amounts to 5.1 emu/g, which corresponds to the SQ value of only 0.38. The squareness parameter serves as a convenient estimate for magnetic hardness of the material, important in adjusting the switching performance of high-speed spintronic devices [45–47]. As such, it provides indirect information about the anisotropic properties of the sample. The fact that even at low temperatures the SQ parameter remains rather low supports our interpretation on the random distribution of local magnetic anisotropy axes between the crystallites.

The coercivity $H_c$ vs $T$ dependence is similar to that of the $M_r$ vs $T$ one, displaying a sharp upturn below ~150 K and reaching the highest value of 57 Oe at 3 K (Fig. 2(d)). The observed high sensitivity of $H_c$ and $M_r$ to the temperature variation in the low-temperature region might reflect the changes in the behavior of magnetic anisotropy [42,48], and this point will be discussed below.

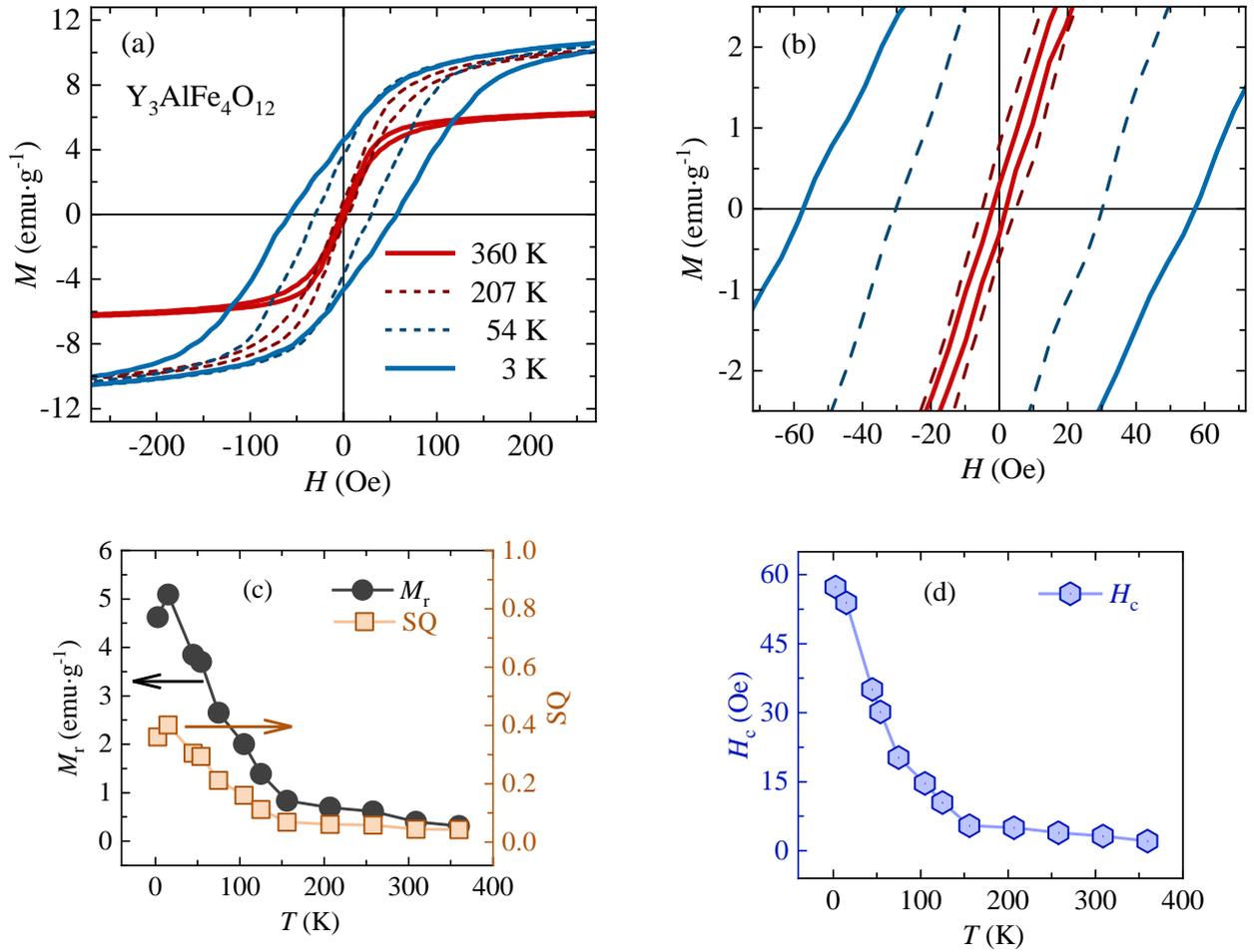

Fig. 2. (a) Magnetic hysteresis loops for the $Y_3AlFe_4O_{12}$ ceramic sample at selected temperatures. (b) Corresponding low-field regions of the hysteresis loops. (c) Temperature dependencies of the remnant magnetization (black) and loop squareness parameter $SQ = M_r/M_s$ (blue), determined from the hysteresis loops in panel (a). (d) Temperature dependence of the coercivity $H_c$ for the $Y_3AlFe_4O_{12}$ ceramics.

## 4. Discussion
### 4.1. Effective anisotropy constant

One way to quantify the magnetic anisotropy from the experimental hysteresis loops is to employ the law of approach to saturation (LAS) [49,50], which has been extensively used in recent studies [51–54]. The main advantage of this approach is that LAS describes a high-field part of the magnetization curve $M(H)$, where the majority of processes related to the domain structure transformation are already finalized, and thus, can be ignored. Changes in magnetization, occurring in this region, are thought to originate exclusively from reversible rotation of local magnetic moments, which, in turn, depend on the magnetic anisotropy of individual crystallites. As such, the LAS model reflects the coherent rotation of spins slightly deviated from the equilibrium orientation, which is dictated by the competition between the local anisotropy field and the externally applied one. In practice, however, additional contributions, which consider structural and magnetic inhomogeneities in real samples, should also be accounted for (see below).

The expression, that we found to provide satisfactory results in our case of bulk YAIG ceramics, is as follows:

$$M = M_s\left(1 - \frac{a_1}{\sqrt{H}} - \frac{a_2}{H} - 0.0762\frac{K_{Eff}^2}{M_s^2 H^2} - 0.0383\frac{K_{Eff}^3}{M_s^3 H^3}\right) + \chi_0 H \qquad (1)$$

where the term with $a_1$ describes the contribution of random anisotropy fluctuations in the sample [55], $a_2$ – contribution of mechanical stresses which arise around dislocations and other imperfections [56], $\chi_0 H$ – paramagnetic-like increase of spontaneous magnetization (paraprocess). The terms that include $K_{Eff}$ are directly related to coherent rotation of spins [48,57]. It should be noted, that $K_{Eff}$ values, determined using LAS, are not only the result of averaging the contributions from magnetocrystalline anisotropy, but also incorporate the magnetoelastic term, which is not described by the LAS model explicitly, i.e. $K_{Eff} = K_1 + \lambda\sigma$ [58], where $K_1$ is the actual first-order magnetocrystalline anisotropy constant, $\lambda\sigma$ – the contribution from the magnetoelastic energy. Therefore, the $K_{Eff}$ parameter, extracted using the Eq. (1), is referred to as the *effective* magnetic anisotropy constant.

As seen from Fig. 3(a), the fitted curves well reproduce the experimental hysteresis loops in the entire temperature range. The extracted saturation magnetization $M_s$ vs $T$ dependence is presented in Fig. 3(b). The $M_s$ value at 3 K reads 12.8 emu/g, which is more than three times lower than that of a pure YIG, reported in literature [59,60] (62 emu/cc as opposed to 196 emu/cc at 3 K) and falls off slowly to 7.44 emu/g at 360 K. Standard error for $M_s$ evaluation was less than 5%. In contrast, the temperature dependence of the effective magnetic anisotropy constant $K_{Eff}$ (see Fig. 3(c)) shows exponential-like behavior, displaying the maximum value $|K_{Eff}| \approx 13.2$ kerg·cm$^{-3}$ at 3 K. The estimated standard error for $K_{Eff}$ turns out to be higher than that for $M_s$, ~30%, which is especially noticeable at low temperatures. At this point it is important to emphasize that the LAS model parameters tend to vary depending on the choice of the high-field region within which the fitting procedure is carrying out [49,55]. This point was also noticed during the calculations in the present study. Whereas $M_s$ experiences only slight variation depending on the fitting field region extension, the $K_{Eff}$ parameter appears much more sensitive, which is characterized by the elevated error values.

By including the terms with $a_1$ and $a_2$ in the model we were able to catch the intermediate fields region (before the higher order terms, $1/H^2$ and $1/H^3$, take in), considerably improving goodness of the resulting fits and providing reliable results on $K_{Eff}(T)$. One should note, however, that both parameters $a_1$ and $a_2$ did not exceed 3 Oe$^{1/2}$ and 20 Oe, respectively, remaining almost independent of temperature.

Fig. 3(c) depicts the temperature dependence of the effective anisotropy field, calculated as $H_a = 2|K_{Eff}|/M_s$. It is noteworthy that at temperatures higher than 150 K, $H_a$ attains rather low values (< 40 Oe) which are smaller than $H_a$ in pure YIG [9,61–63]. In particular, at room temperature $H_a \approx 33$ Oe in YAIG as opposed to 87 Oe in YIG. This means that the strategy to develop materials for spintronic devices with low controlling fields and improved switching performance may consist in doping YIG with non-magnetic ions.

To date, only scarce data are available related to magnetic anisotropy of Al-doped YIG ferrites. As far as we know, the only relevant data were reported in Ref. [64]. It was found that the room-temperature values of $|K_{Eff}|$ were 1.69 kerg·cm$^{-3}$ and 0.43 kerg·cm$^{-3}$ for $Y_3Al_{0.08}Fe_{4.17}O_{12}$ and $Y_3Al_{1.16}Fe_{3.84}O_{12}$, respectively. The value of $|K_{Eff}|$ we obtained for YAIG at room temperature (0.71 kerg·cm$^3$) is in compliance with these data.

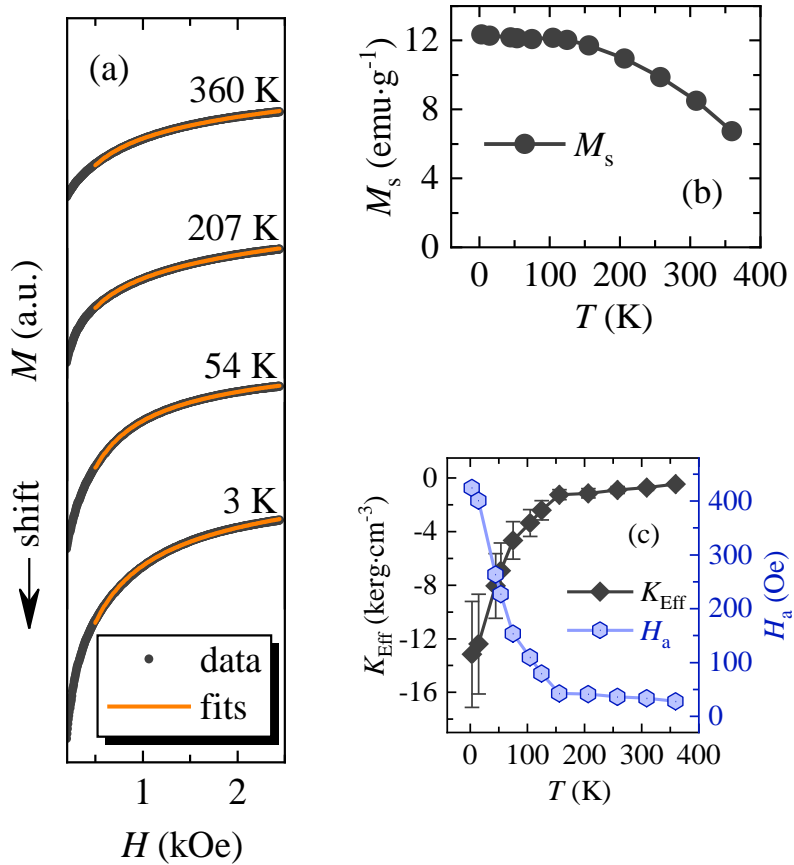

Fig. 3. (a) Calculated high-field $M(H)$ curves shown for selected temperatures and retrieved via fitting the saturation regions of the experimental hysteresis loops with the LAS model (solid lines). Corresponding experimental data are shown as dots. (b) Saturation magnetization $M_s$ vs $T$ curves as deduced from the LAS fitting procedure. (c) Temperature dependencies of effective magnetic anisotropy constant $K_{Eff}$ (black) and respective anisotropy field $H_a = 2|K_{Eff}|/M_s$ (blue), extracted from the LAS model fits.

### 4.2. Spin wave stiffness constant

The saturation magnetization behavior of yttrium iron garnets of the type $Y_3D_nFe_{5-n}O_{12}$ with non-magnetic (diamagnetic) cation substitution $D^{3+}$ for $Fe^{3+}$ can be described within the Brillouin-Weiss theory, as was initially proposed by *L. Néel* [65] and later extended and verified in a number of other works [60,66,67]. Briefly, the essence of the Brillouin-Weiss (BW) approach lies in the determination of the garnet sublattice moments $M_d$ and $M_a$ (the moments of the tetrahedral and octahedral spin sublattices, respectively), exposed to the inter- and intra-sublattice molecular fields, which are represented in the model via the corresponding coefficients $N_{ij}$. Thus, a net magnetic moment $M = M_d + M_a$ for a diamagnetically doped YIG can be deduced by careful considerations of appropriate reductions in the molecular field coefficients $N_{ij}$. *Appendix* provides the details of the application of this approach to the case of $Y_3AlFe_4O_{12}$ ceramics under study.

Fig. 4(a) compares the experimental $M_s$-vs-$T$ dependence for the Al-substituted sample $Y_3AlFe_4O_{12}$ (dots) with the theoretical thermo-magnetization curve (orange solid line), obtained via Brillouin-Weiss (BW) calculations as described in *Appendix*. Black curve corresponds to the BW calculations carried out for a pure $Y_3Fe_5O_{12}$ garnet with the use of the parameters obtained by *G.F. Dionne* [68].

One can observe more than three-fold drop of the zero-temperature total magnetization ($M_{YAIG}(0) = 1.5\ \mu_B$/f.u. as opposed to 5 $\mu_B$/f.u. for pure YIG) as a result of the predominant substitution of Al ions in tetrahedral sites [28,69] (see Fig. A1 in *Appendix*). Another prominent change is the expected decrease of Curie temperature of the YAIG to $T_C \approx 436$ K, directly related to the expected weakening of the exchange interactions upon Al substitution (compare values, provided in Table A1 in *Appendix*).

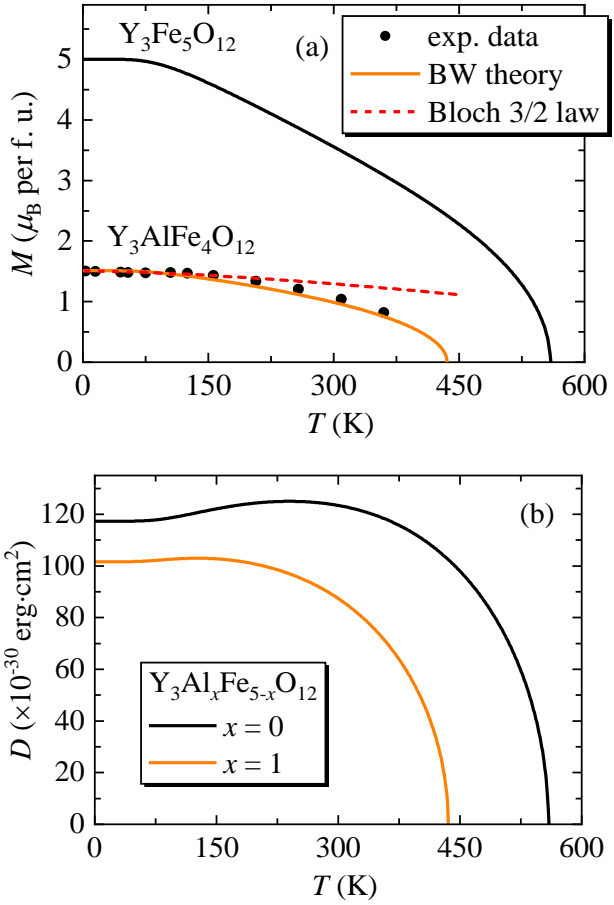

Fig. 4. (a) The experimental $M_s$-vs-$T$ dependence for the Al-substituted sample $Y_3AlFe_4O_{12}$ (dots) and the theoretical thermo-magnetization curve, obtained via Brillouin-Weiss (BW) calculations (orange solid line). Black curve corresponds to BW calculations carried out for a pure $Y_3Fe_5O_{12}$ garnet. Bloch law dependence is shown as dashed line. (b) Temperature dependencies of spin-wave stiffness constant $D$, calculated for Al-substituted garnet ($x = 1$) and pure YIG ($x = 0$).

Extending the BW formalism and its molecular-field approximation with appropriate quantum-mechanical treatment, *C.M. Srivastava* and *R. Aiyar* elaborated the expression for direct determination of the spin-wave stiffness constant $D(T)$ based on the sublattice magnetizations of a garnet [70]:

$$\frac{D(T)}{D(0)} = \frac{(M_d(T)/M_d(0))(M_a(T)/M_a(0))}{(|M_d(T)|-|M_a(T)|)/(|M_d(0)|-|M_a(0)|)},$$

$$D(0) = \frac{5}{16}(5J_{da} - 8J_{aa} - 3J_{dd})a_0^2,$$

(2)

where $D(0)$ – zero-temperature spin-wave stiffness according to *R.L. Douglass* [71], $a_0$ – the lattice constant extracted from the XRD data ($a_0 = 12.321(5)$ Å). $J_{ij}$ are the exchange integrals recalculated from the corresponding molecular field coefficients $N_{ij}$, as suggested by *G.F. Dionne* [68].

It should be pointed out that the presence of the intra-sublattice exchange integrals $J_{dd}$ and $J_{aa}$ in the Eq. (2) yields unrealistic magnitudes of $D(0)$. In fact, even for the exchange constants of an undoped YIG, a negative value is obtained [25]. The authors of [70] explained this due to the fact that spin-wave excitation in YIG is governed by oscillations of the net magnetic moment, i.e. simultaneous deviations of spins in both sublattices $M_d$ and $M_a$, while they remain strictly collinear to each other. In this situation only $J_{da}$ constant should be included in the calculations, with the other parameters set to zero. The resulting spin-wave stiffness dependence $D(T)$, calculated with above considerations, is shown in Fig. 4(b) (orange line) in comparison to the curve for an undoped YIG (black). Both the expected sharp decrease as $T \rightarrow T_C$ and the shift of $T_C$ itself are well recreated in the calculated dependencies. Only about 20 % reduction in $D$ at 0 K is seen as a result of doping the garnet with Al.

As there is no reliable data on low-temperature behavior of spin-wave stiffness $D$ for $Y_3AlFe_4O_{12}$ to date, we decided to check how the calculated thermo-magnetization curve $M(T)$ (Fig. 4(a)) aligns with the Bloch law $M(T) = M(0)\left(1 - \beta T^{3/2}\right)$ [53]. It is known that the $\beta$ parameter in the Bloch law is closely related to the spin-wave stiffness $D(0)$, describing changes in magnetization due to magnon excitations at $T \rightarrow 0$ [72]. Thus, we used the obtained value $D(0) = 102 \times 10^{-30}$ erg·cm² to calculate the $\beta$ parameter ($2.77 \times 10^{-5}$ K$^{2/3}$) and the corresponding temperature dependence of the YAIG magnetization. Despite the expected departure as $T \rightarrow T_C$, the resulting $M(T)$ curve, shown in Fig. 4(a) (dashed), demonstrates good agreement with $M_s$ vs $T$ dependence at low temperatures.

In addition to retrieving the well-resolved $D$-vs-$T$ curve, we also interpolated the temperature dependence of the effective magnetic anisotropy constant $K_{\text{Eff}}(T)$, utilizing the classical Zener's power-law $K_{\text{Eff}}(T) = K_{\text{Eff}}(0) \cdot \left[M(T)/M(0)\right]^{10}$ [73], with the $M(T)$ taken from the BW model fit (orange curve in Fig. 4(a)), and the initial $K_{\text{Eff}}(0)$ value found from the LAS (13.2 kerg/cm³; Fig. 3(c)).

### 4.3. Domain structure parameters

Having the values of the effective magnetic anisotropy constant $K_{\text{Eff}}(T)$ and the spin-wave stiffness $D(T)$ available at hands, we are able to calculate the domain wall width and the critical single-domain size using the well-known expressions, $\delta_w = \pi\sqrt{A_{ex}/|K_1|}$ and $D_{cr} = 72\sqrt{A_{ex}|K_1|}/4\pi M^2$, respectively [74]. The exchange stiffness constant, calculated as $A_{ex} = M \cdot D/(2g\mu_B)$ [42] was close to the experimental values, found for garnets in the literature [75]. Fig. 5 represents the deduced temperature dependencies for $\delta_w$ and $D_{cr}$.

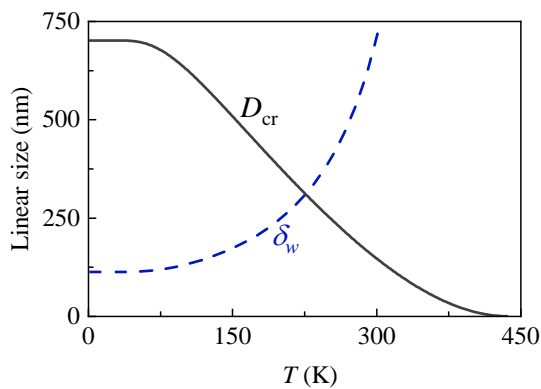

Fig. 5. Temperature dependence of domain wall thickness $\delta_w$ and critical single-domain size $D_{cr}$, obtained for the Y$_3$AlFe$_4$O$_{12}$ sample using effective magnetic anisotropy constant deduced from the LAS (see Fig. 3(c)) and exchange stiffness constant from BW calculations (see Fig. 4(b)).

The critical single domain size $D_{cr}$ exhibits the highest value ~700 nm at 0 K and gradually falls off as the temperature increases, reflecting the competition between the dominant inter-sublattice exchange and enhanced thermal fluctuations. At room temperature $D_{cr}$ reaches values close to those reported for YIG nanoparticles, ~150 nm. The domain wall width $\delta_w(T)$ yields ~110 nm at 0 K, which is also of the order of the values observed in garnet materials [76]. In contrast to $D_{cr}(T)$ dependence the $\delta_w(T)$ one diverges asymptotically when $T \rightarrow T_C$, well recreating the behavior predicted by the theory [77].

### 5. Conclusions

Based on the detailed analysis of the results of quasistatic magnetic measurements, carried out on Y$_3$AlFe$_4$O$_{12}$ ceramic samples synthesized by a unique technological route, the key intrinsic parameters governing the static and dynamic magnetic behavior of Al-doped YIG garnets are extracted, and their field and temperature dependence determined.

It is found that the values of the saturation magnetization, $M_s(T \rightarrow 0 \text{ K}) \approx 12.3$ emu/g ($1.5\,\mu_B$/f.u.), and Curie temperature, $T_C \approx 436$ K, for the samples under investigation agree well with those predicted by our numerical simulations. As temperature is lowered, the coercivity, $H_c$, remnant magnetization, $H_r$, and hysteresis loop squareness, SQ, display a relatively sharp rise below ~150 K, while the saturation magnetization, $M_s$, changes only slightly (<20 %) within 3 – 250 K.

The absolute value of the effective anisotropy constant, $K_{Eff}$, shows an exponential-like growth with decreasing temperature, and reaches a maximum of $|K_{Eff}| \approx 13.2$ kerg·cm$^{-3}$ at $T \approx 0$ K. In the low-temperature region (up to 250 K), the absolute values of $K_{Eff}$ are 2 – 3 times smaller than those reported for pure YIG. Above 150 K, the effective anisotropy field, defined as $H_a = 2|K_{Eff}|/M_s$, is low relative to $H_a$ in pure YIG. In particular, at room temperature $H_a \approx 33$ Oe in YAIG as opposed to 87 Oe in YIG.

Based on the results of our measurements and numerical simulations, we determined the characteristic temperature dependence of the exchange stiffness and spin wave stiffness parameters: $A_{ex}(T)$ and $D(T)$, respectively. Over the whole temperature range studied, the $D(T)$ value for YAIG is less than 80% of that for pure YIG. This implies that a lower energy is needed to excite long-wavelength spin waves in YAIG as compared to YIG.

The critical single domain size, $D_{cr}$, has the highest value of ~700 nm at $T \approx 0$ K and gradually falls off as the temperature increases, reflecting the competition between the dominant inter-sublattice exchange and enhanced thermal fluctuations. At room temperature, $D_{cr}$ reaches values close to the characteristic of YIG nanoparticles, ~150 nm. The domain wall width $\delta_w(T)$ is found to be ~110 nm at $T \approx 0$ K, and diverges asymptotically when $T \rightarrow T_C$, well recreating the behavior predicted by theory.

Understanding the thermal and doping effects on the above key parameters makes it possible to control the static and dynamic magnetic behavior of YIG-based materials and paves the way for developing spintronic devices with low operating fields and improved switching performance.

**CRediT authorship contribution statement**

**Vladyslav Borynskyi**: Formal analysis, Numerical simulation, Writing - Original Draft. **Dariia Popadiuk**: Magnetic measurements, Data analysis. **Anatolii Kravets**: Formal analysis, Magnetic measurements. **Yuliia Shlapa**: Investigation, Resources. **Serhii Solopan**: Investigation, Resources, Methodology. **Vladislav Korenivski**: Supervision, Writing – review & editing. **Anatolii Belous**: Supervision, Project administration. **Alexandr Tovstolytkin**: Conceptualization, Project administration, Writing – review & editing.

**Declaration of Competing Interest**

The authors declare that they have no known competing financial interests or personal relationships that could have appeared to influence the work reported in this paper.

**Data Availability**

Data will be made available on request.

**Acknowledgment**

This work was partially supported by the IEEE Magnetics Society Program "Magnetism for Ukraine – 2023″ (STCU Project No. 9918) and NAS of Ukraine (grant No 0124U002212 in the framework of the Target Program "Grants of the NAS of Ukraine to research laboratories/groups of young scientists of the NAS of Ukraine" (2024–2025)). V. B. is grateful for the support from the NRFU grant # 2020.02/0261. D. P., A. K., and V. K. gratefully acknowledge financial support from Wenner-Gren Foundation (grant GFU2022-0011), Swedish Research Council (VR grant 2018-03526), and Swedish Strategic Research Council (SSF grants UKR22-0050 and UKR24-0002).

**Appendix**

The model developed in [60,66,68] pictures a net magnetic moment of a garnet to be the vector sum of moments of two antiparallel Fe-sublattices, $M = M_d + M_a$ (where $M_d$ and $M_a$ – are the magnetizations of the tetrahedral and octahedral spin sublattices, respectively). Temperature dependencies of the sublattice moments are determined by the corresponding Brillouin functions:

$$M_d(T) = M_d(0) B_{Sd}(h_d),$$
$$M_a(T) = M_a(0) B_{Sa}(h_a),$$
(A1)

where $M_d(0)$ and $M_a(0)$ – magnitudes of $d$- and $a$-sublattice magnetic moments at zero temperature, $h_d$ and $h_a$ – molecular fields acting on the respective sublattices. First important mechanism that determines the sublattice magnetizations $M_d$ and $M_a$ is that spins, both within and between the sublattices, are antiferromagnetically coupled by superexchange, mediated by the $O^{2-}$ anions, which divide the neighboring $d$-$d$, $a$-$a$ and $d$-$a$ site pairs. The Brillouin-Weiss approach models these interactions by introducing the appropriate contributions to the molecular fields in the arguments of the Brillouin functions [66,67]:

$$h_d = \frac{gS\mu_B^2 N_A}{k_B T}(N_{dd} M_d - N_{da} M_a),$$
$$h_a = \frac{gS\mu_B^2 N_A}{k_B T}(-N_{ad} M_d + N_{aa} M_a),$$
(A2)

where $g$ is spectroscopic splitting factor, $S$ – spin state of $Fe^{3+}$ ($S = 5/2$), $\mu_B$ – Bohr magneton, $N_A$ – Avogadro constant, $k_B$ – Boltzman constant, $N_{ii}$ and $N_{ij}$ – molecular field coefficients responsible for exchange interactions within the $i$-th sublattice and between the two, respectively. It is generally assumed that $N_{da} = N_{ad}$, as is also kept during calculations in this work.

Diamagnetic doping of the tetrahedral sublattice, which is predominant for $Al^{3+}$ cations [28,69], reduces the corresponding sublattice magnetization $M_d$ and, thus, the difference with the opposite one $|M_d| - |M_a|$, leading to a significant drop in the net magnetization of a garnet. This effect can be well seen in the $M_d(T)$ and $M_a(T)$ curves (solid lines), calculated for $Y_3AlFe_4O_{12}$ ceramics in the present study, Fig. A1.

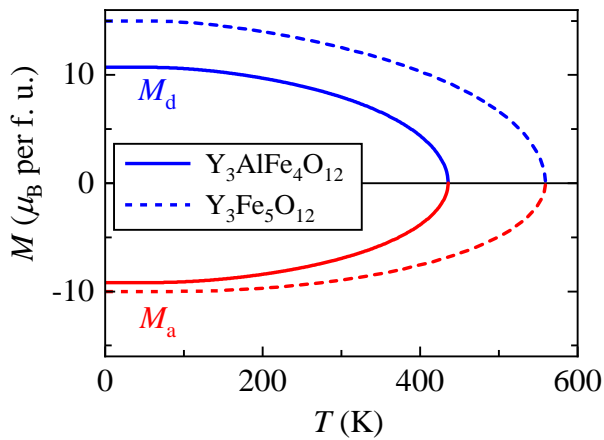

Fig. A1. Calculated temperature dependencies of the $d$- and $a$-sublattice magnetic moments (marked as $M_d$ and $M_a$, respectively), shown for Al-substituted garnet (solid line) and pure YIG (dashed), respectively.

Another important factor, which was successfully addressed by *G.F. Dionne* [66,68], is spin canting, arising due to the substitution of magnetic $Fe^{3+}$ by diamagnetic cations $D^{3+}$, that subsequently causes a reduction in magnetization of a doped garnet.

By analyzing a wide set of empirical data on garnets with various compositions, *G.F. Dionne* found solutions for the molecular field coefficients $N_{ij}$ and the zero-temperature sublattice magnetizations $M_d(0)$ and $M_a(0)$, expressed in terms of normalized substitution levels $x$ (for the $d$-sites substitution) and $y$ (for the $a$-sites) of a garnet $Y_3[D_xFe_{1-x}][D_yFe_{1-y}]O_{12}$. The expressions for $M_d(0)$ and $M_a(0)$, as originally suggested in Ref. [66], are:

$$M_d(0) = 3gS(1-x)(1-0.1y),$$
$$M_a(0) = 2gS(1-y)(1-x^{5.4}),$$
(A3)

It was also shown in this paper that diamagnetic cation substitution in an arbitrary sublattice weakens the exchange between the neighboring spins of the opposite sublattice, reducing not only the corresponding molecular field coefficient $N_{ii}$, but also the $N_{da}$ from the values for an undoped YIG [60]. This continues until the coupling between the sublattices $N_{da}$, which plays a decisive role in maintaining spins within the sublattices parallel to each other, becomes weaker than the $d$-$d$ and $a$-$a$ superexchange, translating the garnet into antiferromagnetic state.

Following the same logic and using the Eqs. (A1-A3), we performed the independent adjustment of the $N_{ii}$, $N_{ij}$, $x$ and $y$ parameters to make the resulting $M$ vs $T$ curve best fit to the experimental data. As shown in Fig. 4(a) (main text), the calculated curve (orange solid line), obtained with the parameters given in the Table A1, well reproduces the temperature dependence $M_s(T)$ (black circles). The obtained molecular field coefficients are in excellent agreement with the values predicted by *G.F. Dionne* [66]. The sublattice substitution levels $x$ and $y$, used in calculating the zero-temperature sublattice magnetizations in Eq. (A3), provide the expected concentration of Al cations $n = 3x + 2y = 1$ (in atoms per formula unit), thus, validating the high level of chemical purity granted by our new proposed precipitation technique [24,41].

| Composition | $x$ | $y$ | $M_d(0)$ ($\mu_B$ f.u.) | $M_a(0)$ ($\mu_B$ f.u.) | $N_{dd}$ (mol/cm$^3$) | $N_{aa}$ (mol/cm$^3$) | $N_{da}$ ($N_{ad}$) (mol/cm$^3$) |
|---|---|---|---|---|---|---|---|
| $Y_3Fe_5O_{12}$ [66] | 0 | 0 | 15 | 10 | –30.4 | –65.0 | 97.0 |
| $Y_3AlFe_4O_{12}$ | 0.28 | 0.08 | 10.72 | 9.21 | –28.34 | –42.07 | 84.79 |

Table. A1. Parameters deduced from the Brillouin-Weiss model by fitting the experimental $M_s(T)$ data.